\documentclass[10pt,journal,compsoc]{IEEEtran}
\ifCLASSOPTIONcompsoc
  \usepackage[nocompress]{cite}
\else
  \usepackage{cite}
\fi
\usepackage{color}
\usepackage{times}
\usepackage{epsfig}
\usepackage{amsmath}
\usepackage{amssymb}
\usepackage{caption}
\usepackage{multirow}
\usepackage{graphicx}

\pdfoutput=1

\ifCLASSINFOpdf
\else
\fi
\begin{document}
%
\title{\emph{Hair-GANs}: Recovering 3D Hair Structure\\ from a Single Image}
%
%
%
%

\author{Meng~Zhang \quad
        Youyi Zheng \quad
\IEEEcompsocitemizethanks{
\IEEEcompsocthanksitem M. Zhang, Y. Zheng are with the State Key Lab of CAD\&CG, Zhejiang University, China.\protect\\
E-mail: ~zyy@cad.zju.edu.cn

}
}

%
%

\markboth{IEEE Transactions on Visualization and Computer Graphics,~Vol.~XX, No.~XX. }
{Shell \MakeLowercase{\textit{et al.}}: Bare Demo of IEEEtran.cls for Computer Society Journals}
%



\IEEEtitleabstractindextext{%
\begin{abstract}
We introduce \emph{Hair-GANs}, an architecture of generative adversarial networks, to recover the 3D hair structure from a single image. The goal of our networks is to build a parametric transformation from 2D hair maps to 3D hair structure. The 3D hair structure is represented as a 3D volumetric field which encodes both the occupancy and the orientation information of the hair strands. {Given a single hair image, we first align it with a bust model and extract a set of 2D maps encoding the hair orientation information in 2D, along with the bust depth map to feed into our Hair-GANs.} With our generator network, we compute the 3D volumetric field as the structure guidance for the final hair synthesis. The modeling results not only resemble the hair in the input image but also possesses many vivid details in other views. The efficacy of our method is demonstrated by using a variety of hairstyles and comparing with the prior art.
\end{abstract}

}

\maketitle

\IEEEdisplaynontitleabstractindextext

%
\IEEEpeerreviewmaketitle

\section{Introduction}\label{sec:introduction}
Techniques of 3D avatar modeling are becoming increasingly popular in today’s emerging VR and AR applications.
However, hair modeling, one of the most crucial tasks, still remains challenging due to
the complexity and variety of hairstyles in the real world. Methods of single-view hair modeling~\cite{chai2016autohair}~\cite{hu2015single}~\cite{chai2015high-quality}
are considered as a much more user-friendly in comparison with those of multi-view modeling methods~\cite{echevarria2014capturing}
~\cite{herrera2012lighting}~\cite{hu2014robust}~\cite{jakob2009capturing}~\cite{luo2013structure}~\cite{paris2008hair},
which usually require specialized equipments in controlled studio environment and long processing cycles.

Due to the lack of information at views distant from the input one, single-view hair modeling techniques
usually rely on a large database containing hundreds or thousands of synthetic hairstyles
used as prior knowledge of hair shape, distribution, or complex structure. These data-driven methods~\cite{chai2016autohair}~\cite{hu2015single}~\cite{hu2017avatar}
come with some problems. Firstly, the requirement of large storage for hair database restricts their application on
resource-constrained platforms such as mobile devices. Second, the quality of modeling results relies on
the retrieved hair exemplars which are searched from the quantity-limited database.
Although post-refinement is introduced to improve the detail accuracy, the structure of the final result is still
confined by the initially retrieved exemplar. In addition, the greedy searching process is
slow and difficult to balance the selection criteria between the local detail similarity
and the global shape similarity.

Recently, deep learning has been intensively exploited in many fields of researches.
One of the most attractive characteristics of deep learning methods is their effectiveness
in converting big datum into high-dimensional feature representations. Those learned features are sufficiently qualified to describe a new data and to set up a space mapping from the input to the target output, totally independently from those training datum. The method of~\cite{chai2016autohair} and the concurrent works of~\cite{zhou2018hairnet} and~\cite{VVA2018Saito}
introduce deep learning into single-view hair modeling. In Chai et al.~\cite{chai2016autohair}'s method, they use CNN networks
to detect hair regions and direction predictors from images as preprocessing for the subsequent hair modeling. Zhou et al.~\cite{zhou2018hairnet}
use an auto-encoder to directly learn the correspondence from 2D orientation map to hair stand assembly parameterized to a 2D scalp map of low resolution as in~\cite{wang2009example}. Saito et al.~\cite{VVA2018Saito} put an effort to achieve end-to-end 3D hair inference by leveraging the latent space of auto-encoders to build a bridge between the 2D image and the 3D hair structure. 

\begin{figure}[t!]
    \centering
    \includegraphics[width=\linewidth]{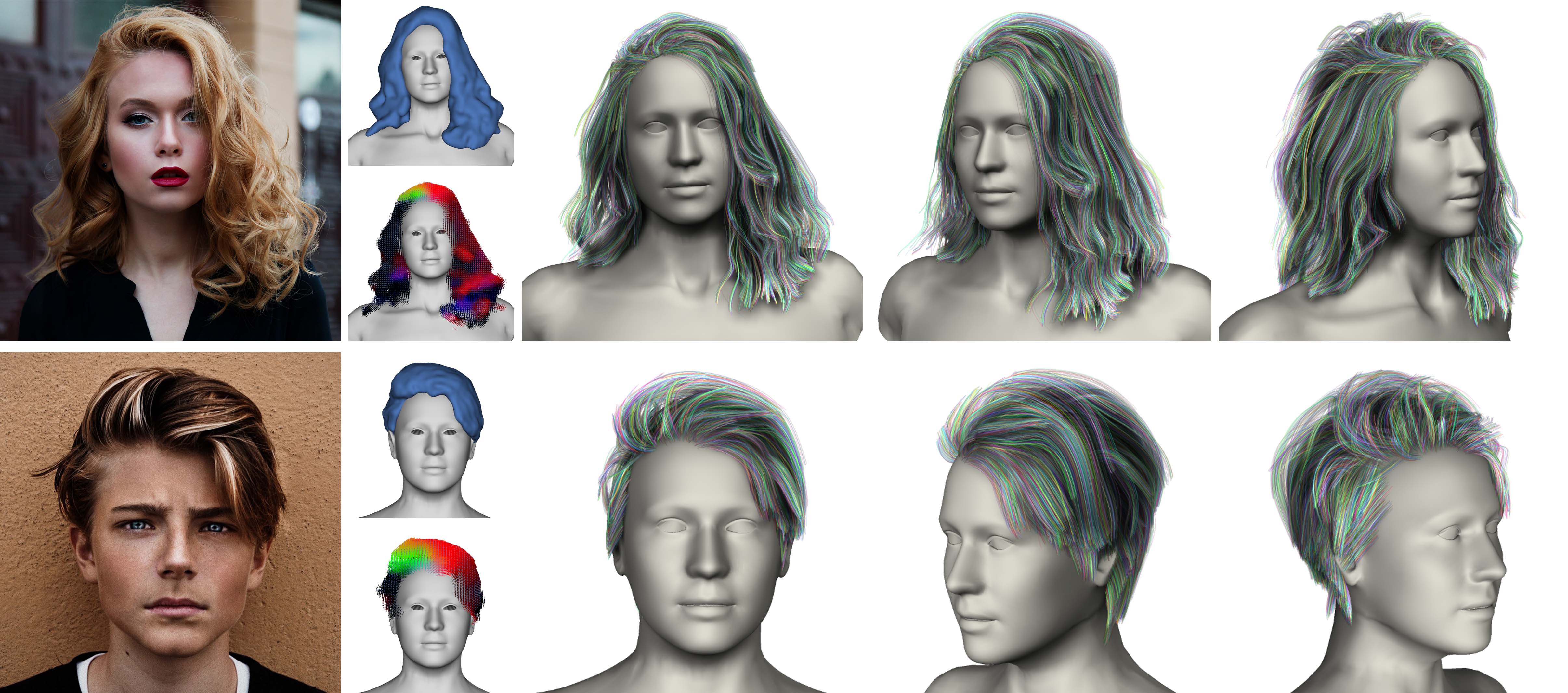}
    \caption{\label{fig:teaser}
    Given single-view inputs of a variety of hairstyles (the first column),
    our generator network recovers 3D hair structures represented as the rough shape and orientation field (the second column) to guide hair synthesis.
    Complete strand-level 3D hair models (the last three columns) are finally generated that resemble the hairs in input images.}
\end{figure}

In this paper, we introduce \emph{Hair-GANs}, an architecture of generative adversarial networks
to recover the 3D hair structure from a single input image. As studied in the vision communities, generative adversarial networks are capable of capturing the model distribution better than those of CNN and auto-encoder models because their adversarial training nature tends to learn more generic distance measures by themselves rather than hand-coded \cite{goodfellow2014generative}. In order to build the data for the network trainings, we use 3D artificially synthetic hair database. Like the deep learning methods of~\cite{zhou2018hairnet} and~\cite{VVA2018Saito}, the large amount of database is
put aside after the training of the network is completed. 

A particular challenge of single-view hair modeling is that the input image only provides 2D information,
lacking the cues along the depth direction in the 3D space.
Therefore, a direct 2D-to-3D training may lead to unconstrained results. We thus involve the bust depth map into the input tensor of our network as a condition to our GANs, which provides depth prior
for hairs growing around the human body. Meanwhile, we convert a succession of 2D features to a single channel of 3D features by a dimensional expansion layer to aggregate 3D knowledge from 2D convolutional neural networks. Inspired by the adversarial structure of GANs~\cite{goodfellow2014generative}, we minimize the objective loss of discriminator to enlarge the difference between the real and the fake, and competitively optimize the generator
to reduce the distance between the output and the ground truth, not only voxel-wisely but also in the latent space of the discriminator. Our method can generate a high-quality 3D strand-level hair model
with a single image as input, guided by the volume of occupancy and orientation field generated by our generator network.
3D hair structure recovered by our \emph{Hair-GANs} takes care of high perceptual-quality details
along the depth direction, rather than a smooth rough hair shape as in~\cite{VVA2018Saito}.
Fig.~\ref{fig:teaser} shows the efficacy of our method for both shot and long, straight and curly hairstyles.

In summary, our contributions are:

\begin{itemize}
    \item We introduce the architecture of GANs for single-view hair modeling. Our GANs transform 2D orientation maps into 3D volumetric field which encodes both the occupancy and orientation information of hair strands;
    \item We propose a dimension expansion layer into the design of our generator network which converts a succession of 2D features to a single channel of 3D features;
    \item We optimize the generator parameters by considering both the output and the latent features of the discriminator. 
\end{itemize}

\section{Related work}\label{sec:Related_work}
Hair is one of the vital components of digital characters in online games, virtual worlds,
and virtual reality applications. The techniques of modeling high-quality 3D hair are extensively studied
in computer graphics, which usually require professional skills and days of laborious manual work.
Please refer to the survey~\cite{ward2007survey} for a detailed discussion of seminal hair modeling methods.

Image-based hair modeling is a promising way to create compelling hair structures from captured hair images.
According to the number of images required, methods of image-based hair modeling can be roughly separated into
multi-view hair modeling and single-view hair modeling. Multi-view hair modeling methods~\cite{echevarria2014capturing}~\cite{herrera2012lighting}~\cite{hu2014robust}~\cite{jakob2009capturing}
~\cite{luo2013structure}~\cite{paris2008hair} create high-quality 3D hair modeling from a number of views,
which often require complex hard-ware setup, well-controlled environment, and long processing cycles.
They are not consumer-friendly since these multi-view capture systems and professional skills are not easily accessible to average users.
While single-view hair modeling methods are becoming increasingly popular and important
as sing-view, un-calibrated images are widely available on the Internet. Chai et al.~\cite{chai2012single-view}~\cite{chai2013dynamic}
first introduce the technique of single-view hair modeling by utilizing different kinds of prior knowledge,
including layer boundary and occlusion,  and shading cues~\cite{chai2015high-quality}.
A major problem of their methods is the lack of control over the geometry at views distant from the input image.

Data-driven hair modeling methods provide a conceptually persuasive prior knowledge of the entire hairstyle,
with 3D synthetic hairstyle database. Hu et al.~\cite{hu2015single} assemble different hairstyles
searched from the database by fitting with a few user strokes to reconstruct a complete hair shape.
Chai et al.~\cite{chai2016autohair} bring forward the model remixing step to precomputation stage.
About 5-40 candidates are found from their enlarged database and
afterward, they perform deformation on these candidates to achieve a model result with detail similarity.
In order to enrich their 3D hair database to 40K models, they cluster hair strands and
recombine those cluster models. Zhang et al.~\cite{Zhang2017fourviews} only use the candidate searched by contour fitting
for their four-view based hair modeling to build a smooth rough hair shape and the style details are introduced
by texture melding and helix-fitting. In~\cite{zhang2018rgbdHair}, they introduce a local patch-based searching strategy
to find candidates with local style patterns enough to guide hair synthesis, instead of to find those with the same style in global.
All these data-driven methods need storage for hundreds or thousands of hairstyle database.

The recent success of deep learning also brings significant improvement in the fields of hair modeling.
Chai et al.~\cite{chai2016autohair} present a fully automated hair modeling method
by replacing user interactions with deep convolutional neural networks to hair segmentation and
hair growth direction estimation. Hu et al.~\cite{hu2017avatar} introduce deep learning based hair attribute classifier
to improve the candidate retrieval performance of the data-driven method. In order to get an end-to-end learning from
2D image knowledge to 3D hair representation, Zhou et al.~\cite{zhou2018hairnet} use encoder-decoder architecture
to generate hair strands represented as sequences of 3D points for 2D orientation fields as input.
But their hair representation is parameterized to a low-resolution grid on the scalp, which leads the modeling result of a low quality.
In a concurrent work, Saito et al.~\cite{VVA2018Saito} demonstrate that 3D occupancy field and the corresponding flow field
with high resolution are easily handled by neural networks and compatible with traditional strand-based representation
for high-fidelity modeling and rendering. However, in their method, the occupancy field and flow field are decoded separately
from the same volumetric latent space. They fine-tune the pre-trained network of ResNet-50 to encode input image to the hair coefficients
which are aligned to the volumetric latent space by their trained embedding networks.
Since there's compression during the processing of encoder and the information loss in latent coefficient alignment, their results lack corresponding details
between the input image and the output hair structure. In comparison, with 2D information maps as input,
our method is more direct to train \emph{Hair-GANs} to predict the 3D volumetric field encoding
both the occupancy and orientation information, with consideration of detail correspondence
between the input image and the modeling result.

Generative adversarial networks (GANs) is introduced by Goodfellow et al.~\cite{goodfellow2014generative}
as a framework to construct a generative model that can mimic the target distribution. The goal of GANs is
to train the generator model by iteratively training the discriminator and generator in turn.
Conditional GANs~\cite{MirzaO14CGANs} are a type of GANs using conditional information for the discriminator and generator,
regarded as a promising tool on image domains, e.g., the conditional image synthesis~\cite{SyntheGANs2017Augu}, the generation of the images from text~\cite{reed2016generative},
and image to image translation~\cite{CycGANs2017Zhu}. We adopt GANs to recover 3D hair structure
from 2D image information, taking advantage of the power of GANs to re-create the distributions of complex data sets.
We make use of the latent space of discriminator to enforce the similarity of the ground truth and target output in distribution. Our \emph{Hair-GANs} aims to learn a parametric transformation from 2D information maps to 3D volumetric occupancy and orientation field with no intermediate latent space.

\section{Overview}\label{sec:Overview}
We first clarify our unified model space where all synthetic hair database and the identical bust model are aligned. Based on the unified model space, we generate the training data of the ground-truth 3D volumetric field coupling with the corresponding 2D orientation and confidence maps (\S\ref{sec:data_preparation}). Next, we introduce the architecture and loss functions of our \emph{Hair-GANs} (\S\ref{sec:hair_gans}).
Similar to the original GANs, our networks are also composed by a discriminator and a generator.
Given a real hair image as input, by using our trained \emph{Hair-Generator}, we can recover
the 3D hair structure based on the 2D orientation and confidence maps, along with the bust depth map (all of which are extracted from the image), and finally
synthesis a high-quality 3D hair model (\S\ref{sec:hair_synthesis}).

\section{Data preparation}\label{sec:data_preparation}
Inspired by \cite{zhang2018rgbdHair}, we regard a hairstyle as a fusion of local style patterns distributed
around human body, in contrast to \cite{chai2016autohair}~\cite{hu2015single}~\cite{zhou2018hairnet}, where different hairstyles are treated as
different combination of styled hair strands. Similar to those previous researches, we collect
an original hair dataset with about 300 3D artificial hair models provided by \cite{chai2016autohair},
which have already been aligned to an identical bust model.
We define a unified model space (\S\ref{subsec:model_space}) to prepare our training data (\S\ref{subsec:training_data})
including both the ground-truth 3D volumetric field $\mathcal{Y}$ along with the 2D hair information maps $\mathcal{X}$.

\subsection{Unified model space}\label{subsec:model_space}
\begin{figure}[t!]
    \centering
    \includegraphics[width=\linewidth]{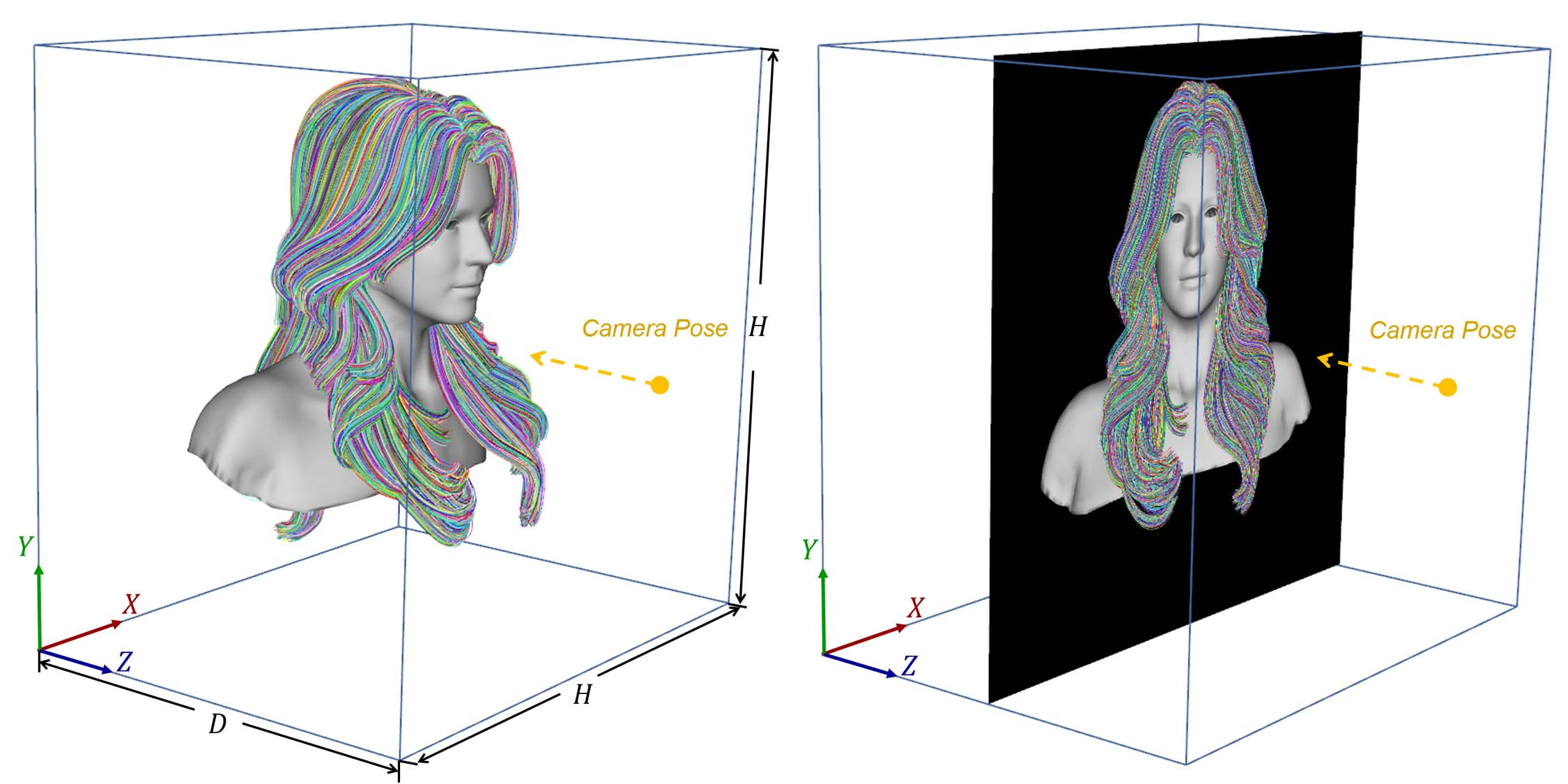}
    \caption{\label{fig:modelSpace}
    Left: A bounding box set up with a camera pose for 3D volume definition and 2D capture.
    Right: An image plane aligned to the bounding box with an image rendered by the defined camera projection.}
\end{figure}

We define a bounding box as the boundary of our model space, where we generate
the ground-truth 3D hair orientation volume and capture 2D hair orientation and confidence maps.
Fig.~\ref{fig:modelSpace} illustrates our defined bounding box and 2D capture as an example.

\emph{Bounding box.} The model space is bounded by a bounding box defined in consideration of the bust model
and all database hair models except some extreme long hairs (selected manually). Then a 3D volume with the resolution of $128 \times 128 \times 96$
is subdivided inside the bounding box ($H \times H \times D$).

\emph{2D capture.} In order to get 2D information maps $\mathcal{X}$ under the defined model space, we put a camera straight forward to the bust model. The center of the image plane coincides with the center of the bounding box. The 2D image is captured by orthogonal projection with a scale of $1024/H$.
Therefore, the size of the captured image is $1024 \times 1024$.

\subsection{Training data}\label{subsec:training_data}
\begin{figure}[t]
    \centering
    \includegraphics[width=\linewidth]{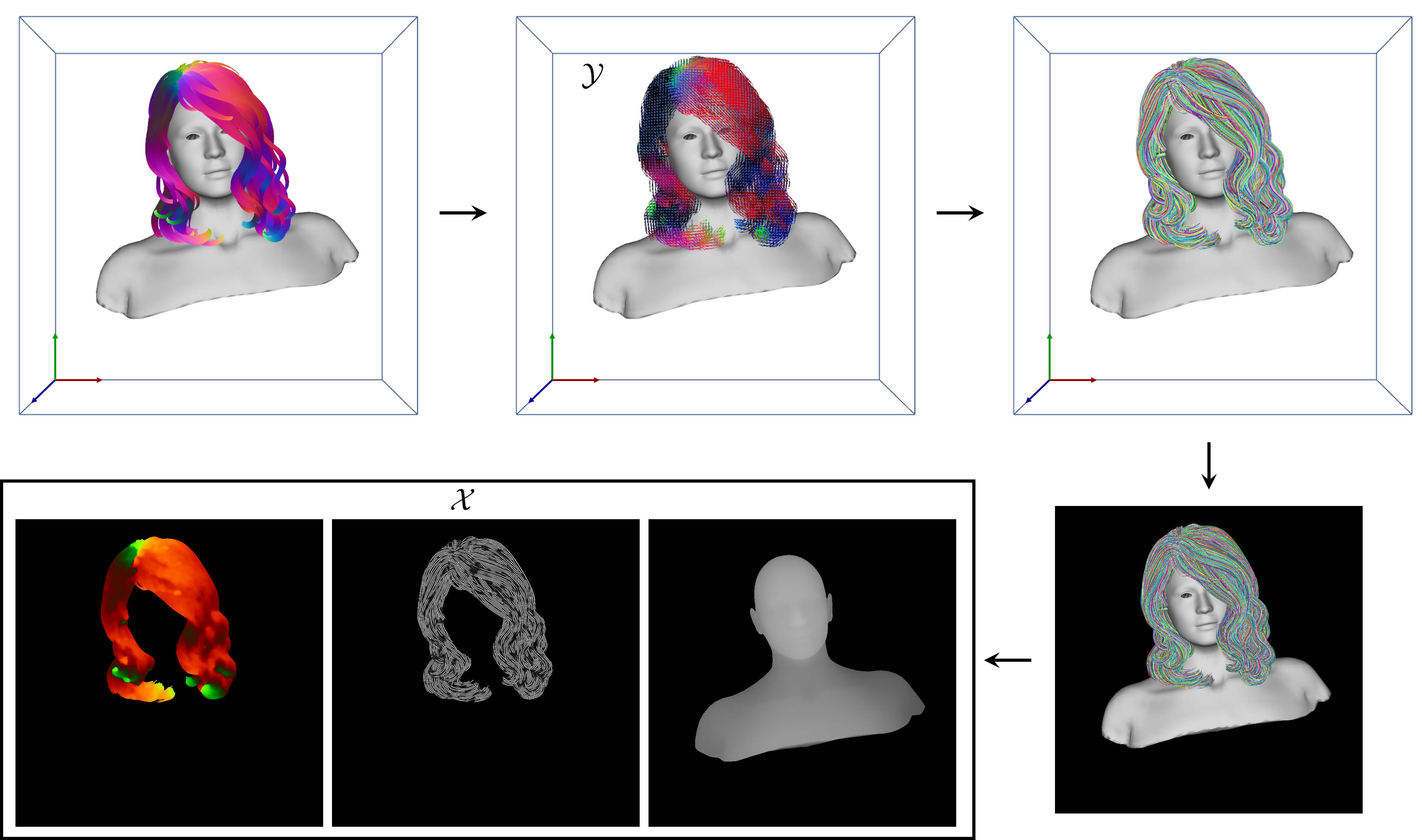}
    \caption{\label{fig:dataPrep}
    Training data generation. The first row from left to right is from a polygon-strip model randomly posed in the defined bounding box to generate the ground-truth volumetric field $\mathcal{Y}$ and to grow hair strands afterwards.
    Then the fourth of the second row is shown a 2D image captured. Based on the 2D image and the bust model,
    $\mathcal{X}$ are calculated made up of orientation, confidence, and bust depth maps (from left to right in the second row).}
\end{figure}
Following \cite{hu2015single}, we double the number of the database by simply flipping each model
and remove the constrained hairstyles such as braids and bounds. There are $303$ hairstyles in our database
varying from short to long, straight to curly. We randomly rotate the hair around the center of the bounding box.
The rotation ranges from $-15^o$ to $15^o$ for X-axle, $-30^o$ to $30^o$ for Y-axle, and $-20^o$ to $20^o$ for Z-axle.
Since all these database models are made up of polygon-strips, same as to Hu et al.~\cite{hu2015single},
we convert the polygon-strips to dense 3D orientation volume regarded as the ground truth $\mathcal{Y}$ and
grow strands afterwards. Then we render the hair strands to the 2D image at the camera view pose defined in \S\ref{subsec:model_space}.
However, in order to remove the difference of real and synthetic hair image, we compute 2D orientation
and confidence maps for the captured image using the iterative method of \cite{chai2012single-view}.
Considering the diverse quality of real images, the iteration for the database randomly ranges from $3$ to $5$. Usually, there is directional ambiguity in the orientation map and Chai et al.~\cite{chai2013dynamic} have confirmed that directional ambiguity should be removed to ensure the correct direction of hair growth. We can project model strand direction to the image plane to update the orientation map to avoid ambiguity. Then we diffuse the orientation with high confidence to obtain the final pixel-dense orientation map and encode direction vectors to color space. In addition, as mentioned earlier, the bust model should also be taken into account as a condition to our networks since hairs are grown from the scalp and distributed around the body. We compute the bust depth map by ray tracing pixel by pixel to get the distance from the bust to the camera, and divide the distance by $D$ to range the value within $[0, 1]$. Finally our network input $\mathcal{X}$ is generated made up of 2D orientation map, confidence map, and bust depth map.

All 2D maps are valued within $[0, 1]$ and both 3D and 2D orientation vectors are encoded in color space.
For each database model, we compute $12$ pairs of $\mathcal{X}$ and $\mathcal{Y}$.
Therefore, we get $3636$ pairs of training datum.
An example of training data generation is shown in Fig.~\ref{fig:dataPrep}

\section{\emph{Hair-GANs}}\label{sec:hair_gans}
With the 2D maps and the bust depth map extracted from the input image, the goal of our Hair-GANs is to produce a 3D orientation volume encoding both the occupancy and orientation information to guide the hair synthesis. The input of the network is a 2D tensor $\mathcal{X}$ with a size of $1024 \times 1024$,
composed of $4$ feature channels that are captured in the unified model space:
hair orientation map (the 2D direction vector $XY$ encoded as the color of RG),
confidence map (the confidence value as the color of gray), and the bust depth map (the depth value as the color of gray). The output is a 3D tensor $\mathcal{Y}$ of size $128 \times 128 \times 96$,
where the hair orientation vectors are encoded in color of RGB. We first describe the loss functions of our adversarial training networks (\S\ref{subsec:objFunction}). Next, we describe the architecture (\S\ref{subsec:architecture}) and the training strategy of our \emph{Hair-GANs} (\S\ref{subsec:training_strategy}).

\subsection{Loss functions}\label{subsec:objFunction}
The GANs~\cite{goodfellow2014generative} is trained in a strategy of competition between two networks:
the generator and the discriminator. We refer to the function form of WGAN-GP~\cite{gulrajani2017improved}
in order to get an easy training. For our cases, the goal is to train a generator $G(\mathcal{X})$
that maps the input 2D tensor to a desired output 3D tensor $\widetilde{\mathcal{Y}}$ : $\widetilde{\mathcal{Y}} = G(\mathcal{X})$.
In the meantime, the discriminator maximizes the Wasserstein-1 distance
between the generator distribution of $G(\mathcal{X})$ and the target distribution of $\mathcal{Y}$
with a conditional latent projection $P(\mathcal{X})$.

\begin{figure*}[t]
    \centering
    \includegraphics[width=\linewidth]{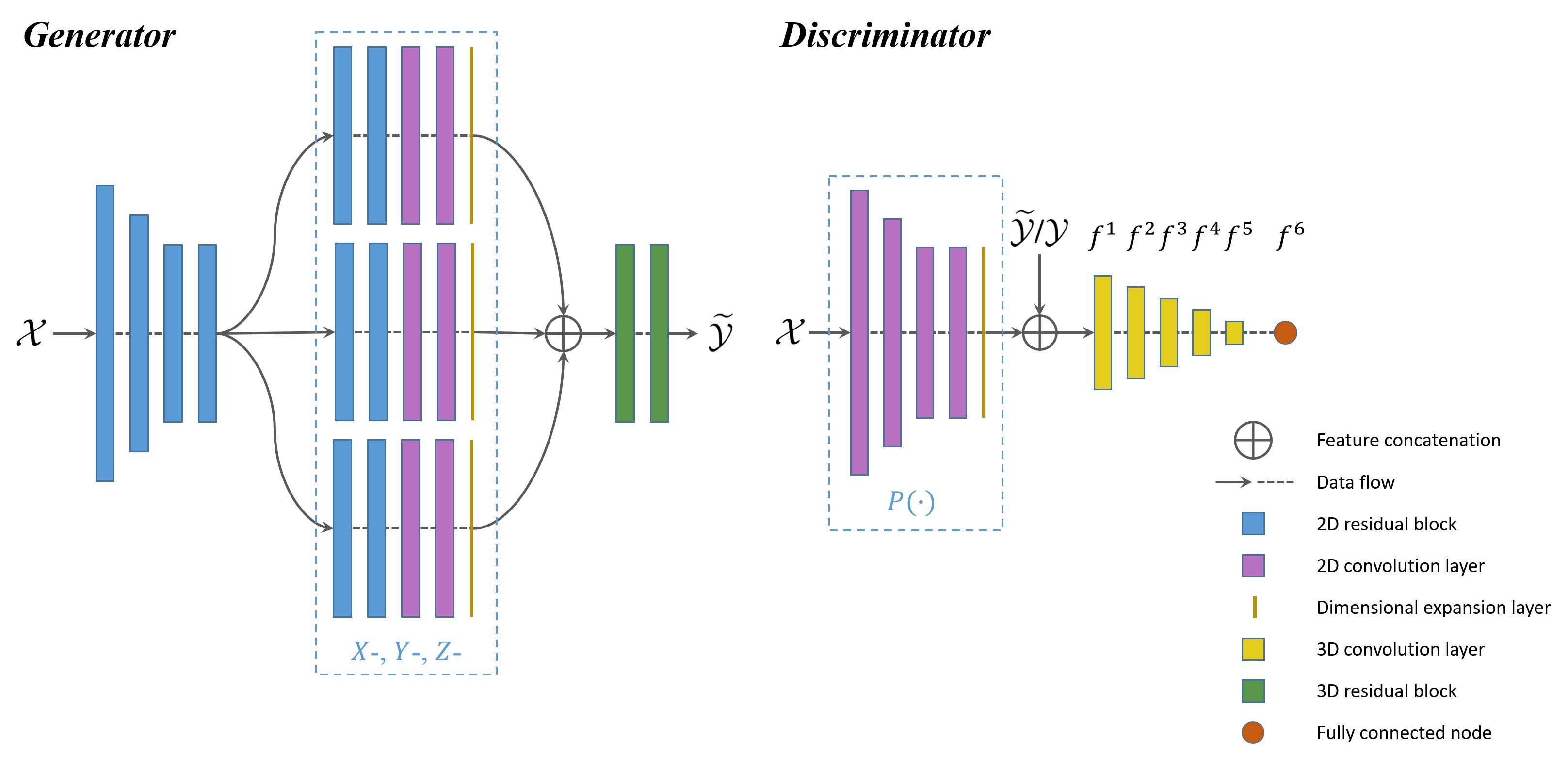}
    \caption{\label{fig:ganArchi}
    An overview of our \emph{Hair-GANs} architecture. The generator and the discriminator are trained in conjunction.}
\end{figure*}

\emph{Discriminator.} The objective of our discriminator is to minimize the energy:
\begin{equation}
   \begin{aligned}
    L_D = &E[D(\widetilde{\mathcal{Y}}, P(\mathcal{X}))] - E[D(\mathcal{Y}, P(\mathcal{X}))] \\
    &+ \lambda E[({\Arrowvert \nabla _{\hat{\mathcal{Y}}}D(\hat{\mathcal{Y}}, P(\mathcal{X}))\Arrowvert}_2 - 1)^2]
    \end{aligned}
\end{equation}
Similar to~\cite{gulrajani2017improved}, the third term in this function is the gradient penalty
for random samples $\hat{\mathcal{Y}}$, that
$\hat{\mathcal{Y}} \leftarrow {\epsilon \mathcal{Y} + (1 - \epsilon) \mathcal{Y}}$,
and $\epsilon$ is a random number in $[0,1]$. The coefficient $\lambda$ is set to 10. $P(\cdot)$ is CNNs
(denoted in Fig.~\ref{fig:ganArchi}) to map 2D tensor $\mathcal{X}$ into a 3D  latent space
in order to be concatenated with $\mathcal{Y}$ or $\hat{\mathcal{Y}}$ referred to the strategy introduced in \cite{reed2016generative}.
And the parameters in $P(\cdot)$ are trained along with those of $D$.

\emph{Generator.} Following the original WGAN-GP~\cite{gulrajani2017improved}, the energy function for generator is
defined as
\begin{equation}\label{eq:naive_ganLoss}
    L_G = -E[D(\widetilde{\mathcal{Y}}, P(\mathcal{X}))]
\end{equation}
However, in our experiment, we find that this function does not work well to optimize the generator,
since the difference of the distribution between the real and the fake cannot be easily determined
by the sign of plus and minus. Inspired by a previous work~\cite{Gatys2016StyleTransfer}
where they use selected layers of pre-trained networks VGG as feature representation to transfer texture style from a source image to a target one, here, we introduce the losses of style and content to our research, where the features are represented in the domains of selected discriminator layers.
Thus, the objective to optimize the generator is to minimize the energy:
\begin{equation}
    \begin{aligned}
        L^*_G &= \alpha L_{content} + \beta L_{style} \\
        &= \alpha \sum_l{L^l_{content}} + \beta \sum_l{L^l_{style}}
    \end{aligned}
\end{equation}
$\alpha$ and $\beta$ are the weighting factors. As described in~\cite{Gatys2016StyleTransfer}, the content loss
is defined by square-error loss between feature representations:
\begin{equation}
    L^l_{content} = \frac{1}{2}\sum_{ik}[f^l_{ik}(\mathcal{Y}, P(\mathcal{X})) -
    f^l_{ik}(\widetilde{\mathcal{Y}}, P(\mathcal{X}))]^2
\end{equation}
Here $l$ is a selected layer, $i$ is the is i-th feature map, $k$ is the index in feature tensor,
and $f$ is the discriminator features (denoted in Fig.~\ref{fig:ganArchi}). The style loss is defined
by the mean-squared distance between the Gram matrices, where each element is calculated
by the inner product between the vectorized feature maps $i$ and $j$:
$A_{ij}^l = \sum_k{f^l_{ik}f^l_{jk}}$. The objective is:
\begin{equation}
    L^l_{style} = \frac{1}{4N^2_l M^2_l}\sum_{ij}[A^l_{ij}(\mathcal{Y}, P(\mathcal{X})) -
    A^l_{ij}(\widetilde{\mathcal{Y}}, P(\mathcal{X}))]^2
\end{equation}
Here $N_l$ is the number of feature maps and $M_l$ is the size of feature tensors
(e.g. if $l=0$, $N_0=3$ and $M_0=128 \times 128 \times 96$).

\subsection{Architecture}\label{subsec:architecture}
Fig.~\ref{fig:ganArchi} and Table~\ref{table:archi} describe the architecture of generator and discriminator in detail. To clearly specify the architecture of our \emph{Hair-GANs}, we use the following notations:
let \emph{in}(resolution, feature channels) and \emph{out}(resolution, feature channels) represent input and output information for operation blocks;
$C$(input channels, output channels, strides) denote the convolutional layer with a ReLU activation followed;
$\varpi$ is dimensional expansion layer, and $\zeta$ is fully connected node.
We use $+$ as the element-wise addition in the residual blocks constituted by $C$, and $I$ denotes the input tensor current layer.
For all 2D convolutional layers $C_2$, the filter size is $5$, and $3$ for all 3D convolutional layers $C_3$.
The operation blocks for $X$-, $Y$-, $Z$-info have the same form of scheme.
\begin{table}
    \begin{center}
        \setlength{\tabcolsep}{3.5mm}
        {
            \begin{tabular}{l}
                \hline
                \multicolumn{1}{c}{\textbf{\emph{Generator}}} \\
                \hline
                \emph{in}$(1024\times{1024},4)\Leftarrow{\mathcal{X}}$\\
                $C_2(4,16,2)+[C_2(4,8,2), C_2(8,16,1)]$ \\
                $C_2(16,64,2)+[C_2(16,32,2),C_2(32,64,1)]$ \\
                $C_2(64,256,1)+[C_2(64,128,2),C_2(128,256,1)]$ \\
                $I+[C_2(256,256,1),C_2(256,256,1)]$ \\
                \emph{out}$(128\times{128},256)$ \\
                \hline
                \multicolumn{1}{c}{$X$-,$Y$-,$Z$-\emph{blocks}}\\
                \emph{in}$(128\times{128},256)$ \\
                $I+[C_2(256,256,1),C_2(256,256,1)]$ \\
                $I+[C_2(256,256,1),C_2(256,256,1)]$ \\
                $C_2(256,128,1)$ \\
                $C_2(128,96,1)$ \\
                $\varpi$ \\
                \emph{out}$(128\times{128}\times{96},1)$ \\
                \hline
                Concat. the \emph{out}s from $X$-,$Y$-,$Z$-\emph{blocks}\\
                \emph{in}$(128\times{128}\times{96},3)$ \\
                $I+[C_3(3,3,1),C_3(3,3,1)]$\\
                $I+[C_3(3,3,1),C_3(3,3,1)]$\\
                \emph{out}$(128\times{128}\times{96},3)\Rightarrow{\widetilde{\mathcal{Y}}}$\\
                \hline
            \end{tabular}
        }
        \setlength{\tabcolsep}{2mm}
        {
            \begin{tabular}{l|l}
                \hline
                \multicolumn{2}{c}{\textbf{\emph{Discriminator}}} \\
                \hline
                \multicolumn{1}{c|}{$P(\cdot)$ \emph{block}} & Concat. $\widetilde{\mathcal{Y}}$/$\mathcal{Y}$ with $P(\mathcal{X})$\\
                \emph{in}$(1024\times{1024},4)\Leftarrow{\mathcal{X}}$ & \emph{in}$(128\times{128}\times{96},4)$\\
                $C_2(4,32,2)$ & $C_3(4,32,2)$\\
                $C_2(32,64,2)$ & $C_3(32,64,2)$\\
                $C_2(64,128,2)$ & $C_3(64,128,2)$\\
                $C_2(128,96,1)$ & $C_3(126,256,2)$\\
                $\varpi$ & $C_3(256,512,2)$\\
                \emph{out}$(128\times{128}\times{96},1)$ & $\zeta$\\
                \hline
            \end{tabular}
        }
    \end{center}
    \caption{\label{table:archi}Architecture of the generator and the discriminator.}
\end{table}

\emph{Generator.} We illustrate the generator as some blocks. The first block with $\mathcal{X}$ as input,
composed of $4$ residual networks~\cite{lim2017enhanced} element-wisely adding activation from earlier layers
to later layers in order to get a residual correction from high- to low-level information,
down-samples feature maps to a latent code from $1024\times{1024}$ to $128\times{128}$, along with the number of
features increasing from $4$ to $256$. Then $X$-, $Y$-, and $Z$-blocks separately encode the latent code
to features with the number of channels as $96$, the resolution along $Z$-axis in the resulting volume.
$\varpi$ converts the succession of 2D features to a single channel of 3D features.
Afterwards, we concatenate the output from $X$-, $Y$-, $Z$-blocks as input into the following 3D residual convolutional networks.
More details please refer to Fig.~\ref{fig:ganArchi} and Table~\ref{table:archi}.

\emph{Discriminator.} Taking into consideration of the correspondence between the 2D input $\mathcal{X}$
and the 3D desired output $\widetilde{\mathcal{Y}}$/$\mathcal{Y}$, inspired by~\cite{MirzaO14CGANs},
we concatenate $\widetilde{\mathcal{Y}}$/$\mathcal{Y}$ with $P(\mathcal{X})$, a feature map encoding $\mathcal{X}$ to
a 3D latent space with the same resolution as $\widetilde{\mathcal{Y}}$/$\mathcal{Y}$.
Then the concatenated 3D feature tensor is convoluted by a number of filters until the layer of $\zeta$ to finally distinguish the real and fake.

\subsection{Training strategy}\label{subsec:training_strategy}
In~\cite{gulrajani2017improved}, the generator updates after five-times training of the discriminator, which takes a lot of time. For time efficiency, we apply the two-timescale update rule (TTUR)~\cite{heusel2017TTUR} to update the discriminator only once instead of five times. We employ the commonly used ADAM optimizer~\cite{kingma2014adam} with $\beta_1=0$
and $\beta_2=0.9$ for training. The learning rate for the discriminator is set to $0.0003$, and $0.0001$ for the generator.
Generally, our \emph{Hair-GANs} are designed to generate a $128\times{128}\times{96}$ 3D volume encoded in
both of the occupancy and orientation fields, using 2D maps as input with a size of $1024\times{1024}$.
The batch size for training is set to $4$.

For the generator objective, the style and content weighting factors are set as: $\alpha=1e-2$, $\beta/\alpha=5e+2$.
Selected layers for content loss are $0,3,6$ and $l= 0,1,2,3,4$ for style loss. Specifically, when $l=0$, $P(\mathcal{X})$ can be removed
from $L^0_{content}$ and $L^0_{style}$.

\section{Hair synthesis}\label{sec:hair_synthesis}
\begin{figure}[b!]
    \centering
    \includegraphics[width=\linewidth]{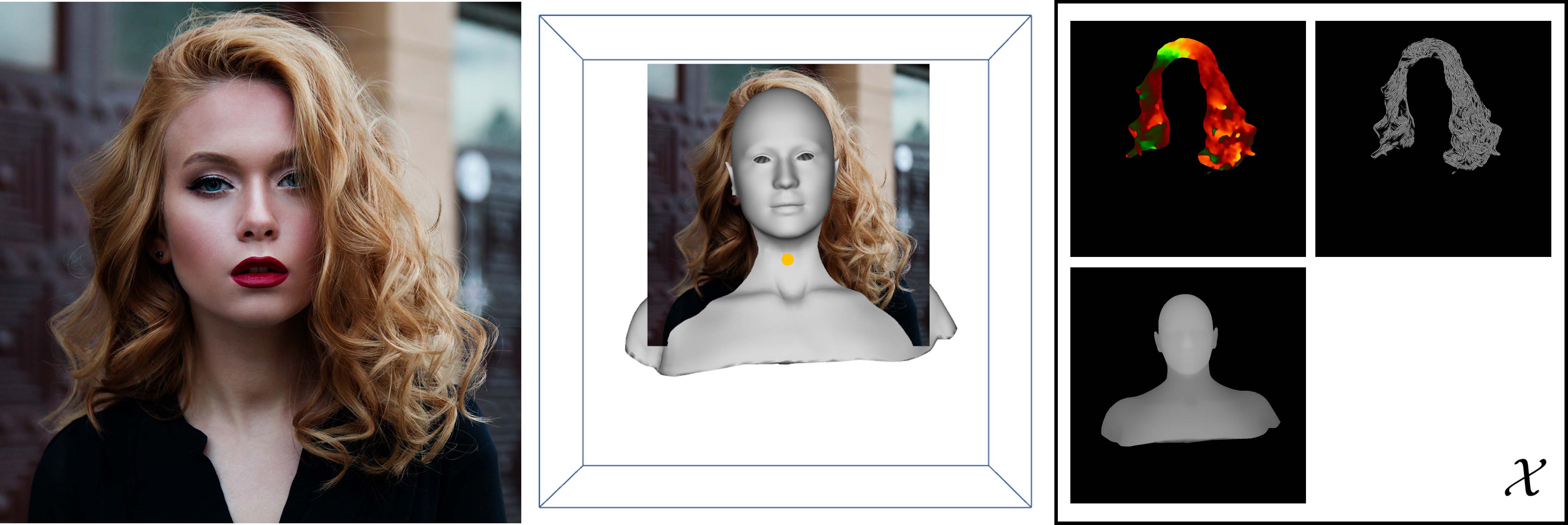}
    \caption{\label{fig:preproc}
    From left to right: input image, image aligned with the bust model fitted,
    and 2D maps as input for the generator network.}
\end{figure}

Given an input image, we first align it to our unified model space and generate 2D information maps and depth map (\S\ref{subsec:preproc})
as input to our generator network. After the 3D orientation volume is generated, we build a rough shape
to confine hair synthesis and finally generate our strand-level hair model (\S\ref{subsec:post-proc}).
\subsection{Preprocessing}\label{subsec:preproc}
Similar to previous single-view methods \cite{chai2016autohair}~\cite{hu2015single}~\cite{chai2012single-view},
with an input image, we first run the face alignment algorithm \cite{cao2014face} to fit the identical bust model to the facial landmark points detected in the image. Also, we segment the hair region and to generate the hair direction predictor to remove directional ambiguity for the orientation map computation afterwards \cite{chai2016autohair}~\cite{chai2013dynamic}. Specifically, assume that we have transformations: $s$ (scaling) $r$ (rotation) and $t$ (translation) involved in the bust fitting, $s$ and $t$ are applied to the image and $r$ is applied to the identical bust model (for correct depth map generation). After the alignment, we capture the image and hair mask by the projection described in \S\ref{subsec:model_space}
and generate maps of orientation, the confidence of orientation, and bust depth referred to \S\ref{subsec:training_data}. We also remove the directional ambiguity with the help of direction predictor from the orientation map. Fig.~\ref{fig:preproc} shows an example of preprocessing.

\subsection{Post-processing}\label{subsec:post-proc} We find that a direct trace along the 3D orientation field computed by our generator network could lead to non-smooth results for some complex hairstyles because the orientation is not smooth in the flow field. Therefore, we apply the marching cube algorithm~\cite{matchingCubes87} on the occupancy field to compute a rough shape and smooth it by Poisson method~\cite{Poisson06M}. We then smooth the exterior orientation field
tangent to the surface of the rough shape. We also optionally refine the orientation field
by warping the image orientation map to the rough shape surface. Constrained by the normal direction of the scalp and the image warping orientation field, we grow hair strands from the scalp uniformly distributed inside the rough shape following the previous method of hair synthesis~\cite{chai2013dynamic}~\cite{hu2015single}. After that, in order to refine the local detail matching with the input image, we run the strands deformation according to the projected image orientation map, as in~\cite{hu2015single}. 
\section{Results and Discussion}\label{sec:rst_discuss}
\begin{figure}[t]
    \centering
    \includegraphics[width=\linewidth]{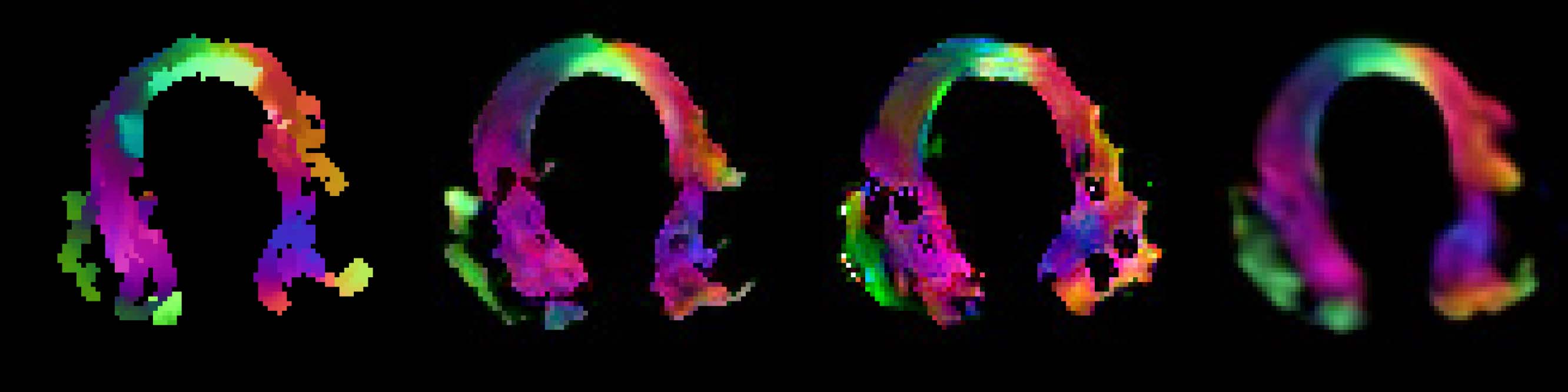}
    \caption{\label{fig:lossComp}
    Slices of volumetric fields. From left to right: the ground truth, our \emph{Hair-GANs} output,
    the result learned using Eq.~\ref{eq:naive_ganLoss} as generator objective, a CNN generator output with only naive $L^0_{content}$ loss (no GAN training).}
\end{figure}

\begin{figure}[!b]
    \centering
    \includegraphics[width=\linewidth]{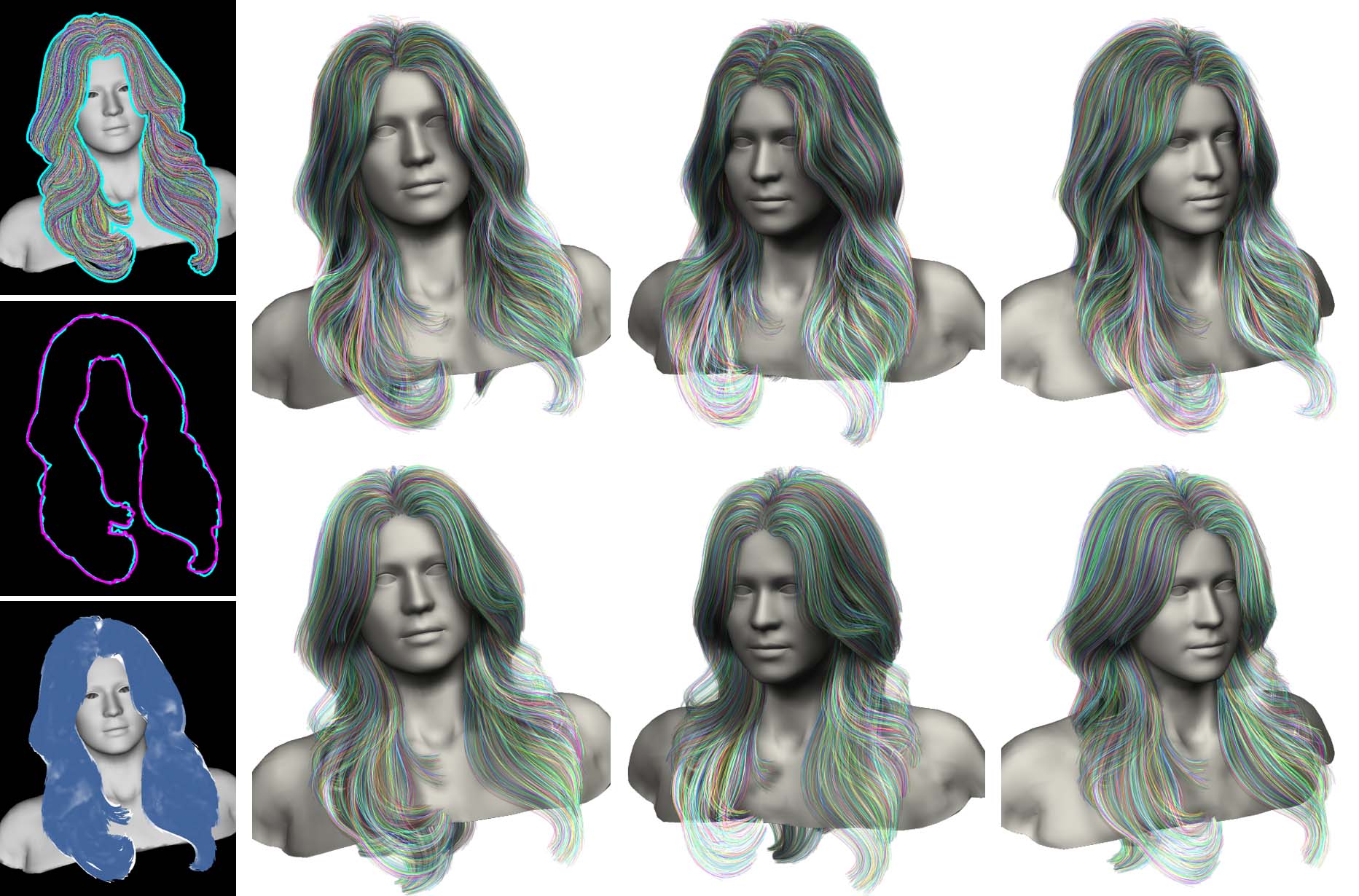}
    \caption{\label{fig:GT_Comp}
    Up to down in the first column: a synthetic image with blue-colored hair contour, comparison of hair mask
    (the blue colored is the mask of the input image, while the magenta colored is mask projected by the modeling result), orientation difference map (white color indicates there's a big difference between the input and the modeling result in orientation map).
    The right three columns: the first row is our hair modeling results and the second row is the ground-truth.}
\end{figure}

\begin{figure}[t]
    \centering
    \includegraphics[width=\linewidth]{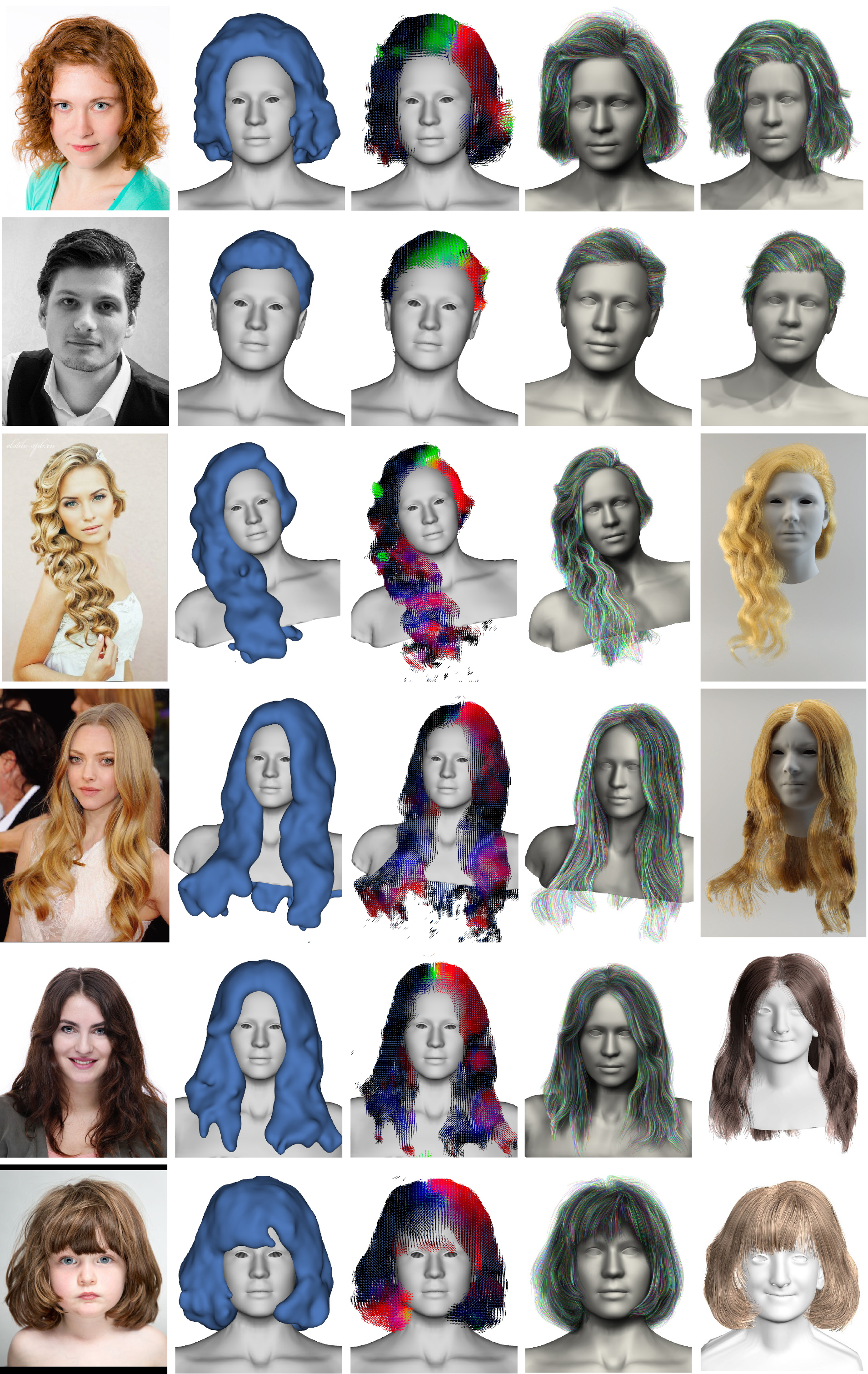}

    \caption{\label{fig:comp}
    Comparisons with \cite{chai2016autohair} (the top two rows), \cite{VVA2018Saito} (the middle two rows) and \cite{zhou2018hairnet} (the bottom two rows).
    For each comparison, from left to right: input image, our results (rough shape, orientation field, 3D hair models), the result of the previous method 
    .}
\end{figure}

\begin{figure}[t]
    \centering
    \includegraphics[width=\linewidth]{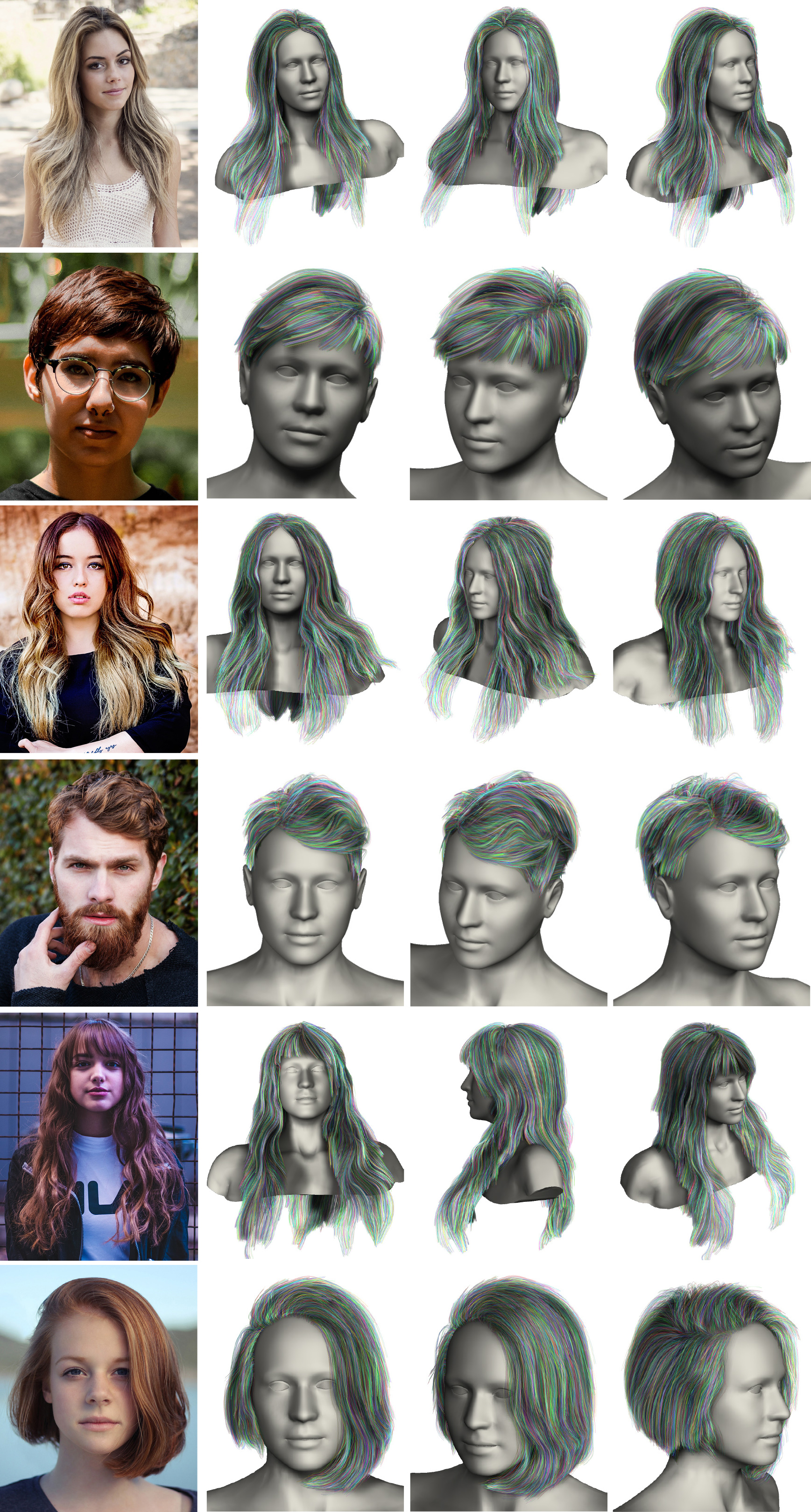}

    \caption{\label{fig:rst}
    Hair modeling results using our \emph{Hair-GANs} with single-view images as input.}
\end{figure}

The implementation of our hair structure recovery runs on a PC with an Intel Core i7-4790 CPU,
32G of memory and an NVIDIA GeForce GTX 1080Ti. It took about 10 days to train our \emph{Hair-GANs}
with 200k generator iterations using our GPU. We use \emph{tensorflow} frameworks to set up our deep learning networks.
With a single-view input image as input, {using our unoptimized pipeline}, it takes less than {$3$ seconds for data preparation by the automatic method of hair parsing~\cite{chai2016autohair}}, about $1$ second to generate 3D volume field using our networks (after tenserflow python call), and less than $30$ seconds to synthesis final hair model.
Fig.~\ref{fig:rst} demonstrates the effectiveness of our method on a variety of hairstyles.

To evaluate the efficacy of our generator objective function, we compare our volumetric filed output
with the ground truth in Fig.~\ref{fig:lossComp}. Additionally, we also train our generator network separately
by using the original function of WGAN-GP (Eq.~\ref{eq:naive_ganLoss}) and the naïve $L_2$ loss.
The original function leads to much ambiguity and noise in both the fields of occupancy and orientation (Fig.~\ref{fig:rst}, middle right).
A naïve training of the generator just using the $L_2$ loss ($L^0_{content}$) results in a blurred output (Fig.~\ref{fig:rst}, right),
just like its usual effect on the super-resolution field~\cite{xie2018tempoGAN}.
Our objective function of the generator considers the distribution similarity of selected discriminator layers in both style and content, which leads to a higher perceptual quality.

We also evaluate the quality of our modeling results in Fig.~\ref{fig:GT_Comp}.
With a synthetic image as input, our \emph{Hair-GANs} produces the modeling results similar to
the ground-truth model even in views distant from the input one. The projected mask of our result can be well matched with the input mask.
Moreover, the projected orientation map is similar to the input image in most of the hair region.

In addition, We compare our method with the state-of-the-art methods of~\cite{chai2016autohair}~\cite{VVA2018Saito}~\cite{zhou2018hairnet} in Fig.~\ref{fig:comp}. The results of Chai et al.~\cite{chai2016autohair} (Fig.~\ref{fig:comp}, top two rows) rely on their database models
which are used as the prior structures of hair modeling. Our results are generated by deep learning networks.
Using our \emph{Hair-GANs}, we can generate more faithful modeling results of hairstyles even
they do not exist in our training data. Saito et al.~\cite{VVA2018Saito} (Fig.~\ref{fig:comp}, middle two rows) use embedding network to
align the principle latent code encoded from the input image to a latent space represented for the 3D field.
There is an unavoidable loss of information during the process of image encoding. Their results do not achieve a good matching of either the hair contour or the orientation field.
Our method makes full use of hair textures from the input image to generate hair structure along the depth direction. Our modeling results more closely resemble both the hair contour and texture in the single-view input images. Zhou et al.~\cite{zhou2018hairnet} (Fig.~\ref{fig:comp}, bottom two rows) compute a feature map of low resolution parameterized on the scalp. Our results show more detail in hair structure.

\section{Conclusion}\label{sec:Conclusion}
We have presented our \emph{Hair-GANs}, an architecture of networks to learn 3D hair structure
from a single-view input image. We use 2D CNNs to learn 3D cues and convert the succession of
2D features to a 3D feature map of a single channel by the dimensional expansion layer.
The core of our method is to follow the training strategy of GANs,
the competition between the generator and the discriminator. In order to produce a volumetric field of high perceptual quality,
we design the generator objective function in consideration of the distribution similarity in all selected discriminator lays,
inspired by the method of style-transfer. Our results on a variety hairstyles resemble the input images
in both of the hair contour and hair texture. The network can also learn more details in the views
distant from the input one.

Our architecture of \emph{Hair-GANs} is subject to a number of limitations, which may inspire interesting future work.
First, like almost all single-view hair modeling methods, our training datum only take the frontal view image into account. Our method would fail when hair is partially occluded.
Second, although our network can produce a considerable volumetric field for complex hairstyles,
hair tracing method still merits some improvement on the quality of the final modeling result.
Current efforts \cite{chai2016autohair}~\cite{hu2014robust}~\cite{zhang2018rgbdHair} improve the hair strand geometry by using the guidance of database models. We hope in the future, there’s a method independent of database to synthesis complex hairstyles, e.g., even the highly curly. Third, we hope to expand our training data dynamically to cover the ever-changing fashion of hairstyles. In addition, the resolution and size of the volumetric field have an impact on the quality of the final modeling results. Some complex hairstyles may need a larger and more subtilized volume. We believe the advances in efficient data structures for 3D-GANs could leverage this problem.


%

\appendices


\ifCLASSOPTIONcompsoc
  \section*{Acknowledgments}
\else
  \section*{Acknowledgment}
\fi


\ifCLASSOPTIONcaptionsoff
  \newpage
\fi



%

\bibliographystyle{IEEEtran}
\bibliography{Hair_GAN}

\end{document}